\documentclass[conference]{IEEEtran}

\usepackage[utf8]{inputenc}
\usepackage{graphicx}
\usepackage{xcolor}
\usepackage{url}

\usepackage{breakurl}
\usepackage{makecell}
\usepackage{enumitem}

\newcommand{\para}[1]{\noindent\textbf{#1}.}
\newcommand{\parait}[1]{\vspace{0.5mm}\noindent\textit{#1}.}

\usepackage{booktabs} 
\newcommand{\ra}[1]{\renewcommand{\arraystretch}{#1}}

\DeclareRobustCommand{\mychar}[1]{%
  \begingroup\normalfont
  \includegraphics[height=\fontcharht\font`\B]{#1}%
  \endgroup
}

\providecommand{\UsePackageFor}[2]{ \ifx#2\undefined\usepackage{#1}\fi }

\UsePackageFor{amssymb}{\boxtimes}
\usepackage[normalem]{ulem}		
\usepackage{bbding}				
\usepackage{endnotes}			
\usepackage{hyperref}			
\usepackage{hyperendnotes}	
\usepackage{pdfcolfoot}			

\ifdefined\OrigFootnote\relax\else
	\newenvironment{FootnoteContent}{}{}
	\let\OrigFootnote\footnote
	\let\OrigFootnoteText\footnotetext
	\renewcommand{\footnotetext}[1]{\OrigFootnoteText{\begin{FootnoteContent}#1\end{FootnoteContent}}}
	\renewcommand{\footnote    }[1]{\OrigFootnote    {\begin{FootnoteContent}#1\end{FootnoteContent}}}
\fi

\definecolor{PurplePlum}{rgb}{0.1,0,0.55} 
\definecolor{Brown}{rgb}{0.5,.25,0}
\definecolor{Orange}{rgb}{1,.3,0}
\definecolor{Gray}{rgb}{.7,.7,.7}
\definecolor{DarkGreen}{rgb}{.1,.41,.1}

\newif\ifBleck
\newcommand\Bleck {\Blecktrue} 

\newcommand\Colour[1] {\color{#1}}

\newcommand\PrintToCLinks{	
  {\Colour{blue}\mbox{
    \hyperlink{w1619}{\sf$\rightarrow$~top}\quad
    \hyperlink{w1031}{\sf$\rightarrow$~toc}\quad
    \hyperlink{w1148}{\sf$\rightarrow$~lof}\quad
    \hyperlink{GreenRoom}{\sf$\rightarrow$~gr}\quad
    \hyperlink{EndNotes}{\sf$\rightarrow$~en}\quad
    \hyperlink{Sargasso}{\sf$\rightarrow$~sg}\quad
    \hyperlink{Index}{\sf$\rightarrow$~idx}
  }}
}
\makeatletter 
\newcommand\ToCLinks{
  \ifx\@onlypreamble\@notprerr		
    \hypertarget{w1619}{}			
  \else
    \AtBeginDocument{\hypertarget{w1619}{}}	
  \fi

  \ifBleck\else	
    \ifdefined\cofoot
      \cofoot{\PrintToCLinks}
      \cefoot{\PrintToCLinks}
    \else
      \def\@oddfoot{\PrintToCLinks}
      \def\@evenfoot{\PrintToCLinks}
    \fi
 \fi
}
\makeatother

\newif\ifEndNotes 

\newcommand\FnSym{{\scriptsize\PencilLeftDown\kern.1em}}		
\newcommand\EnSym {{$\bigtriangledown$}}

\def\MarkupsHowto{} 
\newcommand{\MarkupsHowtoAdd}[1]{\expandafter\def\expandafter\MarkupsHowto\expandafter{\MarkupsHowto{}#1}} 

\newif\ifMarkupsHowtoPrinted 
\newif\ifSuppress 

\newcommand\MakeMarkups[3][.]{

     \Suppressfalse
     \ifBleck\Suppresstrue\fi
     \ifx0#1\Suppresstrue\fi
     \ifx1#1\Suppressfalse\fi
     
     \expandafter\providecommand\csname#2x\endcsname {} 
     \ifSuppress\expandafter\renewcommand\csname#2x\endcsname{\relax}\else
                       \expandafter\renewcommand\csname#2x\endcsname{#3}\fi
                       
     \expandafter\providecommand\csname#2\endcsname {} 
     \ifSuppress\expandafter\renewcommand\csname#2\endcsname[1]{##1}\else
                       \expandafter\renewcommand\csname#2\endcsname[1]{{\csname#2x\endcsname##1}}\fi

     \expandafter\providecommand\csname#2d\endcsname {} 
     \ifSuppress\expandafter\renewcommand\csname#2d\endcsname[1]{\relax}\else
                       \expandafter\renewcommand\csname#2d\endcsname[1]{{\csname#2x\endcsname\sout{##1}}}\fi
                       
     \expandafter\providecommand\csname#2r\endcsname {} 
     \ifSuppress\expandafter\renewcommand\csname#2r\endcsname[2]{{##2}}\else
                       \expandafter\renewcommand\csname#2r\endcsname[2]{\csname#2d\endcsname{##1} \csname#2\endcsname{##2}}\fi

     \expandafter\providecommand\csname#2i\endcsname {} 
     \ifSuppress\expandafter\renewcommand\csname#2i\endcsname[1]{\relax}\else
                       \expandafter\renewcommand\csname#2i\endcsname[1]{\csname#2\endcsname{##1}}\fi

     \expandafter\providecommand\csname#2t\endcsname {} 
     \ifSuppress\expandafter\renewcommand\csname#2t\endcsname[1]{\relax}\else
                       \expandafter\renewcommand\csname#2t\endcsname[1]{{\csname#2x\endcsname{\mbox{$\langle\!\langle$}##1{\csname#2x\endcsname\mbox{$\rangle\!\rangle$}}}}}\fi 

     \expandafter\providecommand\csname#2b\endcsname {} 
     \ifSuppress\expandafter\renewcommand\csname#2b\endcsname[1][empty]{\relax}\else 
                       \expandafter\renewcommand\csname#2b\endcsname[1][\empty]{\ifx\empty##1\empty
                       	\label{#2-bookmark} 
                              \marginpar [\raggedleft\csname#2\endcsname{{\footnotesize\fbox{#2 working here}}~$\Longrightarrow$}]
                                                {\csname#2\endcsname{$\Longleftarrow$~{\footnotesize\fbox{#2 working here}}}}
                       \else 
                       	\marginpar [\raggedleft\csname#2\endcsname{\ifx\empty##1\empty\else\fbox{\tiny\parbox{8em}{\raggedright##1}}~\fi$\Longrightarrow$}]
                                                {\csname#2\endcsname{$\Longleftarrow$\ifx\empty##1\empty\else~{\tiny\fbox{\parbox{8em}{\raggedright##1}}}\fi}}\fi}\fi

     \expandafter\providecommand\csname#2TD\endcsname {} 
     \ifSuppress\expandafter\renewcommand\csname#2TD\endcsname{\relax}\else
                       \expandafter\renewcommand\csname#2TD\endcsname{\csname#2\endcsname{\fbox{#2 to do}}}\fi

     \expandafter\providecommand\csname#2Bar\endcsname {} 
     \ifSuppress\expandafter\renewcommand\csname#2Bar\endcsname{\relax}\else
                       \expandafter\renewcommand\csname#2Bar\endcsname{\csname#2\endcsname{\scriptsize\XSolidBrush}}\fi

     \expandafter\providecommand\csname#2f\endcsname {} 
     \ifSuppress\expandafter\renewcommand\csname#2f\endcsname[2][]{\relax}\else
      \expandafter\renewcommand\csname#2f\endcsname[2][\empty]{ 
        {\mbox{\csname#2x\endcsname\tiny$\boxtimes$}\marginpar{\hsize1cm\csname#2x\endcsname\fbox{\FnSym\footnotemark}}\relax 
        \footnotetext{\csname#2x\endcsname##2}}}\fi

     \expandafter\providecommand\csname#2e\endcsname {}
     \ifSuppress\expandafter\renewcommand\csname#2e\endcsname[1]{\relax}\else%
      \expandafter\renewcommand\csname#2e\endcsname[1]{%
       \global\EndNotestrue
       \mbox{\scriptsize\csname#2x\endcsname$\boxtimes$}\relax%
       \marginpar{\hsize1cm\csname#2x\endcsname\fbox{\EnSym\endnotemark%
                          \hypertarget{ENmark\thepage-\theendnote}{}~\hyperlink{ENtext\thepage-\theendnote}{{\Colour{blue}$\downarrow$}}}%
       }%
       {
        \def\zz{\noexpand#3}%
        \edef\z{~{[Endnote \theendnote\ %
        on p.\noexpand\hypertarget{ENtext\thepage-\theendnote}{}\thepage%
                    ~\noexpand\hyperlink{ENmark\thepage-\theendnote}{{\noexpand\Colour{blue}$\uparrow$}}]}%
        }%
        \expandafter\endnotetext\expandafter{\z\vspace{2ex}\\ ##1\newpage}%
       }
      }\fi

     \expandafter\providecommand\csname#2n\endcsname {}
     \ifSuppress\expandafter\renewcommand\csname#2n\endcsname[1]{\relax}\else%
      \expandafter\renewcommand\csname#2n\endcsname[1]{%
       \global\EndNotestrue
    \marginpar{{\tiny\endnotemark}\hypertarget{ENmark\thepage-\theendnote}{}~\hyperlink{ENtext\thepage-\theendnote}{}}
       {
        \def\zz{\noexpand#3}%
        \edef\z{~{\zz[Endnote (deferred) 
        from p.\noexpand\hypertarget{ENtext\thepage-\theendnote}{}\thepage%
        ]}%
        }%
        \expandafter\endnotetext\expandafter{\z\vspace{2ex}\\ ##1\newpage}%
       }
      }\fi

     \expandafter\providecommand\csname#2fe\endcsname {} 
     \ifSuppress\expandafter\renewcommand\csname#2fe\endcsname[2][]{\relax}\else 
      \expandafter\renewcommand\csname#2fe\endcsname[2][]{ 
       \def\File{##1}\relax
       \ifx\File\empty\csname#2f\endcsname{##2}\else 
        \global\EndNotestrue 
        \mbox{\scriptsize\csname#2x\endcsname$\boxtimes$}
        \marginpar{\csname#2x\endcsname\fbox{\FnSym\footnotemark}}\relax
        \footnotetext{~\csname#2x\endcsname##2\
                             --- See [\EnSym\endnotemark\hypertarget{ENmark\thepage-\theendnote}{}
                             \kern-.2em\hyperlink{ENtext\thepage-\theendnote}{{\Colour{blue}$\downarrow$}}].}\relax
       { 
         \def\zz{\noexpand#3}
         \edef\z{~{\zz[Endnote~\thefootnote~on~p.\noexpand\hypertarget{ENtext\thepage-\theendnote}{}\thepage
                     ~\noexpand\hyperlink{ENmark\thepage-\theendnote}
                     {{\noexpand\Colour{blue}\kern-0.1em$\uparrow$}]}}
                     {\noexpand\footnotesize\noexpand\newline\noexpand\hspace*{2em} (~from file {\noexpand\tt\File.tex}~)}
         }    
         \expandafter\endnotetext\expandafter{\z~\par\input{##1}\newpage}
        } 
       \fi 
      } 
     \fi 

     \ifSuppress\relax\else\ifBleck\relax\else
      \MarkupsHowtoAdd{\par\csname#2t\endcsname{
       $\backslash$\texttt{#2}$\cdots$\ markups are in \textbf{this} colour\ifx#1..\else\ifx1#1.\else, e.g.\ for #1.\fi\fi
       \ifMarkupsHowtoPrinted\relax\else 
        \global\MarkupsHowtoPrintedtrue 
        \begin{quote}\begin{tabular}{l@{\hspace{2em}}p{.7\linewidth}}
         \multicolumn{2}{l}{\texttt{$\backslash$MakeMarkups\ifx#1.\relax\else[#1]\fi\{#2\}\{{\it$\langle$colour command\/$\rangle$}\}}
         				 --- Defines the macros below:}\\
             & see comments at \texttt{$\backslash$MakeMarkups} definition. \\[1ex]
         \texttt{$\backslash$#2\{$\langle$text$\rangle$\}} & Sets \texttt{$\langle$text$\rangle$} in \texttt{#2}'s colour. \\
         \texttt{$\backslash$#2x} & Changes to \texttt{#2}'s colour (until end of context). \\
         \texttt{$\backslash$#2d\{$\langle$text$\rangle$\}} & Sets \texttt{$\langle$text$\rangle$} in \texttt{#2}'s colour with a strikethrough (i.e.\ delete). \\
         \texttt{$\backslash$#2r\{$\langle$this$\rangle$\}\{$\langle$that$\rangle$\}} &
          Strikes through \texttt{$\langle$this$\rangle$} and inserts \texttt{$\langle$that$\rangle$} (i.e.\ replace). \\
         \texttt{$\backslash$#2f\{$\langle$text$\rangle$\}} & Meta-comment: puts \texttt{$\langle$text$\rangle$} in a \texttt{#2}-footnote with a {\tiny$\boxtimes$} in the main text. \\
         \texttt{$\backslash$#2t\{$\langle$text$\rangle$\}} & Use for meta when  \texttt{$\backslash$#2f} isn't allowed (``Not in outer-par mode.'') \\
         \texttt{$\backslash$#2b[$\langle$optional$\rangle$]} & Marginal pointer, with label for hyper-linking directly there. \\
         \texttt{$\backslash$#2e\{$\langle$text$\rangle$\}} & Puts \texttt{$\langle$text$\rangle$} in a \texttt{#2}-endnote with a (big) $\boxtimes$ in the main text. \\[.5ex]
         \texttt{$\backslash$#2n\{$\langle$text$\rangle$\}} & Like \texttt{$\backslash$#2e}
         except there's no reference from the main text. Good for ``decluttering''
         when you still want to have the footnote- or endnote texts as reminders. \\[.5ex]
         \texttt{$\backslash$#2fe[$\langle$this$\rangle$]\{$\langle$that$\rangle$\}} & Makes a \texttt{$\backslash$#2f\{$\langle$that$\rangle$\}} that refers to a \\
           & \texttt{$\backslash$#2e\{$\langle$contents of file this.tex$\rangle$\}}. \\ 
           & Without the optional argument, acts as \texttt{$\backslash$#2f\{$\langle$that$\rangle$\}}. \\[.5ex]
         \texttt{$\backslash$#2Bar} & Inserts ``burn after reading'' symbol \csname#2Bar\endcsname, meaning
          \begin{quote}\begin{itemize}\setlength\itemsep{0pt}
           \item If yours is the only \csname#2Bar\endcsname\ in this (presumably someone else's) footnote, and you are happy that the footnote has been addressed,
           go ahead and comment-out the whole footnote. (The \csname#2Bar\endcsname\ is their request for you to ``approve and remove''.)
           \item If you are not happy, delete only your \csname#2Bar\endcsname\ and follow-on in the footnote
            (in your colour, i.e.\ with \texttt{$\backslash$#2x}) saying why you are not happy.
           \item If you are happy, but there are others' burn-after-reading symbols as well as yours, just delete yours; the other people have not yet responded.
          \end{itemize}
          \end{quote}
          The idea is that when everyone's happy, the last person will comment-out the meta-text. \\[0.5ex]
         \texttt{$\backslash$#2TD} & Inserts {\csname#2TD\endcsname}\ . \\
        \end{tabular}\end{quote}
       \fi
      }}
     \fi\fi
}

\newif\ifNoGreenRoom
\newcommand\MakeGreenRoom {\ifBleck\relax\else\ifNoGreenRoom\relax\else
\newcommand\NewGRLabel[1] {\OldGRLabel{GreenRoom-##1}} 
 \newcommand\NewGRRef[1] 
 {\expandafter\ifx\csname r@GreenRoom-##1\endcsname\relax\OldGRRef{##1}\else\OldGRRef{GreenRoom-##1}\fi}
 \let\OldGRLabel\label \let\label\NewGRLabel
 \let\OldGRRef\ref \let\ref\NewGRRef
 \hrule
 ~\\\begin{center}\Huge \hypertarget{GreenRoom}{Green Room}
 \end{center}~\\
 \hrule
\fi\fi}

\newcommand\EndGreenRoom  {\ifBleck\relax\else\ifNoGreenRoom\relax\else
\let\label\OldGRLabel
\let\ref\OldGRRef
\fi\fi}

\newif\ifNoEndNotes

\newif\ifNoSargasso
\newcommand\MakeSargasso {
 \hypertarget{Sargasso}{}
 \newcommand\NewLabel[1] {\OldLabel{Sargasso-##1}} 
 \newcommand\NewRef[1] 
 {\expandafter\ifx\csname r@Sargasso-##1\endcsname\relax\OldRef{##1}\else\OldRef{Sargasso-##1}\fi}
 \let\OldLabel\label \let\label\NewLabel
 \let\OldRef\ref \let\ref\NewRef
\ifBleck\end{document}\else\ifNoSargasso
\relax
\else
  \hrule
  ~\\\begin{center}\Huge Sargasso
  \end{center}~\\
  \hrule
 \fi\fi
}

\newcommand\EndSargasso  {\ifBleck\relax\else\ifNoSargasso\relax\else
\let\label\OldLabel
\let\ref\OldRef
\fi\fi}

\newcommand\EndDocument {\ifBleck\end{document}\fi} 

\newcommand\Cite[2][\empty] {{\Colour{red}\ifx#1\empty[#2]\else[#2,~#1]\fi}}

\Bleck		
\MakeMarkups[Betty]{B}{\Colour{Brown}}
\MakeMarkups[Seda]{S}{\Colour{DarkGreen}}
\MakeMarkups[Carmela]{C}{\Colour{Orange}}

\usepackage{wasysym} 

\newcommand{\grayCIRCLE}{\textcolor{gray}{\CIRCLE}}

\usepackage{footnote}
\makesavenoteenv{tabular}

\begin{document}

\title{Privacy Engineering Meets Software Engineering.\\
On the Challenges of Engineering Privacy By Design
}

\author{

\IEEEauthorblockN{Blagovesta Kostova}
\IEEEauthorblockA{EPFL\\
blagovesta.pirelli@epfl.ch}

\and
\IEEEauthorblockN{Seda G\"urses}
\IEEEauthorblockA{
TU Delft / KU Leuven \\
f.s.gurses@tudelft.nl}

\and
\IEEEauthorblockN{Carmela Troncoso}
\IEEEauthorblockA{EPFL\\
carmela.troncoso@epfl.ch}
}

\maketitle

\begin{abstract}
Current day software development relies heavily on the use of service architectures and on agile iterative development methods to design, implement, and deploy systems. These practices result in systems made up of multiple services that introduce new data flows and evolving designs that escape the control of a single designer. Academic privacy engineering literature typically abstracts away such conditions of software production in order to achieve generalizable results. Yet, through a systematic study of the literature, we show that proposed solutions inevitably make assumptions about software architectures, development methods and scope of designer control that are misaligned with current practices. These misalignments are likely to pose an obstacle to operationalizing privacy engineering solutions in the wild. Specifically, we identify important limitations in the approaches that researchers take to design and evaluate privacy enhancing technologies which ripple to proposals for privacy engineering methodologies. Based on our analysis, we delineate research and actions needed to re-align research with practice, changes that serve a  precondition for the operationalization of academic privacy results in common software engineering practices.
\end{abstract}




\section{Introduction}

The rise of data-driven services brought with it a wave of
consciousness about their impact on privacy. 
This is reflected in the strengthening of 
legal frameworks for privacy protection~\cite{gdpr},
and in the increased efforts to develop standards focused on building privacy-preserving systems~\cite{ISO27550}.
Academic circles have echoed this trend with an unprecedented number
of papers on privacy technologies at prominent security, privacy, and systems conferences, and with satellite events on privacy engineering at comparable venues~\cite{AlamoKSBM19,KissnerC19}. 

%
Academic work on privacy technology and engineering efforts have multiplied. 
However, they are difficult to translate to current-day software systems.
While some of the translation challenges are due to the lack of economic, legal, and political incentives, we argue that they are also technical.
In this paper, we show that across the board, current privacy research is not attuned to the widespread use of service architectures in combination with agile methods~\cite{GursesVH18,KaldrackM2015}. 
Even when organizations and developers are willing to operationalize privacy research, there is a misalignment between proposed solutions and development practices that is difficult to overcome for both privacy researchers and software engineers.
To illustrate the misalignment, let us consider the problem 
of privacy-preserving billing of pay-per-use services, e.g., usage of roads \cite{BalaschRTPVG10}, electricity \cite{RialD11}, 
or public transportation~\cite{KerschbaumLG13}, whereby a provider can bill clients depending on usage
without learning fine-grained patterns. 

\begin{figure}[t!]
    \centering
	\includegraphics[width=\linewidth]{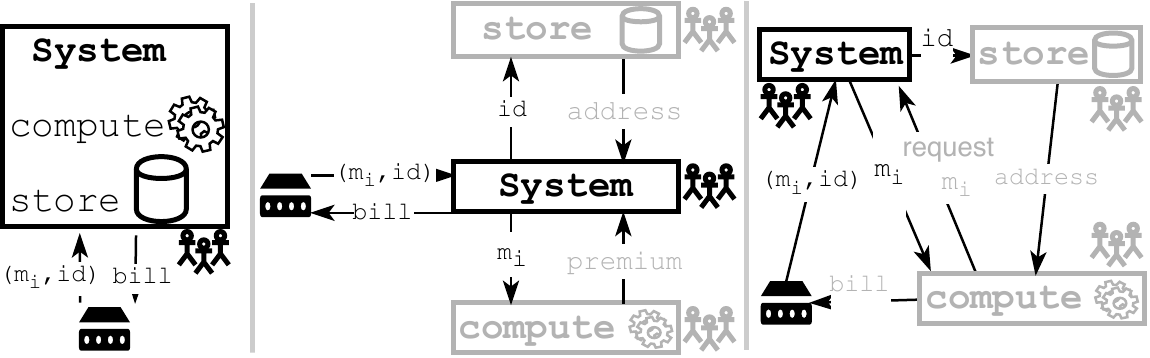}
	\vspace{-7mm}
	\caption{Privacy-preserving billing: research view vs. software practice. Legend: \mychar{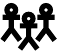} is a development team. \mychar{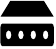} is the device that communicates with the \texttt{System}, \texttt{m}$_i$ is the measurements from the device, \texttt{id} is the id to identify the device. \textcolor{gray}{Grey} elements are outside of the control of the main dev team, whereas black elements indicates what is under their control.}
	\label{fig: example}    
\end{figure}

When modeling, privacy literature typically abstracts the system to a set of users, whose devices record usage measurements,
and a provider that takes as input the measurements and computes a bill that is then sent to the users for payment 
using information stored in a client database.
This system model is shown in Figure~\ref{fig: example}, left.
This model assumes that: 
(i) there is a single development team, 
and (ii) both the privacy-preserving computation engine (\texttt{compute}) and the client database (\texttt{store}) 
are implemented \emph{in-house} by a service provider, i.e., under the control of a \emph{single entity}.

However, under the current reality of software engineering, the system implementation is likely to be very different. 
First, the provider development team is unlikely to implement the service as a software monolith. 
Instead, the service provider typically
uses (third-party) services to cut on costs, such as development, in-house expertise, or maintenance cost.
In our example, the storage of client data, as well as the bill computations would be outsourced to \emph{specialized services}
(see Figure~\ref{fig: example}, center and right). 
Each of these services, and the code used to compose them, 
are developed \emph{independently} by different teams. 
Once the APIs are settled, the service implementations can also evolve without coordination.

This change in the architecture and in the development teams radically changes the environment in which 
privacy engineering, and privacy technologies, must operate.
Among others, the introduction of services introduces new data flows and entities out of the control
of the privacy engineer, which impacts the system threat model that designers can assume.
The shift from sequentially executed waterfall development processes
towards iterative agile methodologies in which system goals and implementations continuously change to address new requirements further complicates the settings of privacy efforts. 
This `moving target' poses a challenge for privacy designs and engineering frameworks that typically assume static system specifications, and rarely evaluate the impact that changes in the system have on the privacy guarantees their solutions provide.

We study this misalignment by conducting a systematic analysis of literature on privacy technologies and privacy engineering methodologies. 
We identify and formulate the points of friction between privacy engineering methodologies and privacy technologies that stem from researchers' implicit conceptions of software engineering and current software development practices.
We find that privacy researchers \textit{systematically} treat engineering as a black box, consequently failing to address these challenges, and how these challenges impact the feasibility of operationalizing solutions coming out of their research. 

A problem with these blindspots is that they cast challenges that arise out of development practices as problems to be solved by software developers. 
This neglects that engineering activities and software production environments raise significant challenges to the adequacy, feasibility, and potential deployment of any privacy research proposal: 
if the available know-how does not address these challenges, neither can software developers.
Hence, studying the problems that arise at the intersection between privacy research and software engineering is of substantial value for the potential impact and of research interest for the questions that are open to the field. 

We end with a discussion that provides a broader view of how we got here and on possible ways forward, laying down a research agenda informed by the conducted analysis. 
We consider legitimate differences but also avoidable gaps between computer science and engineering and across sub-disciplines of computer science, and how these percolate to privacy engineering standards making. 
We also turn the question around and ask how software production may have to change, if we want systems that respect people's privacy. 
In fact, we find that for the wide deployment of privacy-preserving systems, regulation, standards, software industry and software engineering practices also need to evolve to account for the challenges we identify. 
We conclude by identifying low-hanging fruits that could help bring privacy research closer to practice.

\section{The Agile and Services Shift}
\label{sec: agile}

Innovation, economic incentives, new tools, as well as the spread of software to new domains continuously changes the software industry and the way in which software is produced. 
We focus on two major shifts that have shaped software engineering over the last three decades: namely, the shifts (1) from waterfall to agile methodologies in the development of software, and (2) from monolithic to service architectures in the design of software. 
The two have come to redefine the industry, its economic models, but also its relationship to (personal) data, risk, and responsibility~\cite{GursesVH18}.

\subsection{Services}
\label{sec:saas}

Services enable independent development, deployment and evolution~\cite{Zimmermann17}.
Service-oriented architectures make it easier for applications to use services to connect and cooperate within or outside an enterprise.
Application providers can rapidly extend functionality by integrating additional services and can also offer their services to external applications. 
Services are seen as ``a means of delivering value to customers by facilitating outcomes customers want to achieve without the ownership of specific costs and risks''~\cite{ITIL2011}.
This belief in services being low-risk and high-value is paradigmatic for the service-oriented world.
Thus, businesses can respond to market and environmental changes without owning costs and risks -- in other words, services offer \emph{business agility}. 
We can trace this to Conway's law that states that the architecture of IT systems reflects the structure of the organization that designs the system~\cite{Conway68}. Hence, organizational structure and software architecture co-evolve: independent teams produce \emph{specialized} independent services.

A service is an overloaded term. In this paper, we use the service-oriented computing definition:
services are ``self-describing, platform-agnostic computational elements that support rapid, low-cost composition of distributed applications''~\cite{Papazoglou03}.
The service provider publishes a description of their services (an Application Programming Interface, API specification) and a service consumer discovers and calls the services through their API~\cite{Papazoglou03}. 
The service provider and the service consumer are two independent actors. 
Services can be arbitrarily small, including concepts
such as microservices~\cite{DragoniGLMMMS17}, or serverless computing \cite{BaldiniCCCFIMMRSS17}. 

Software applications codify certain logic that operates on data created within or outside the application.
Traditionally, applications have held the execution logic (what function is called when in the form of instructions) and the data flows (what data go where) in one place. 
This type of software architecture, or code organization, is now called a monolithic architecture.
Even though monolithic applications may also use modularization (logical organization of the code in smaller independent packages, or modules), services introduce an additional degree of separation and specialization of the modules into distributed and independent computational elements deployed outside of the boundary and control of the main application.
Service-oriented architectures not only split logically the application execution logic and the data flows but also distribute them across a computational infrastructure into services and a service composition layer~\cite{Papazoglou03,HuhnsS05,PapazoglouH07}. 
The service composition layer is responsible for composition, monitoring, conformance, and quality-of-service of services.
The two established ways for composition of services into an application are orchestration and choreography~\cite{CernyDT17},
which determine how much of the execution logic and data flows the services have.
The service composition layer can keep all the logic and orchestrate the use of different services (e.g., an enterprise service bus)~\cite{Papazoglou03} as in Figure~\ref{fig: example}, center, where the \texttt{System} sends and requests data from \texttt{store} and \texttt{compute} but fully controls the execution flow; or the logic is split among `smart' services that coordinate between themselves, in Figure~\ref{fig: example}, right, the \texttt{System} dispatches requests to the \texttt{store} and  \texttt{compute}, who also connect with each other and the client without intervention from the \texttt{system}.


\begin{figure}[b!]
    \centering
	\includegraphics[width=\linewidth]{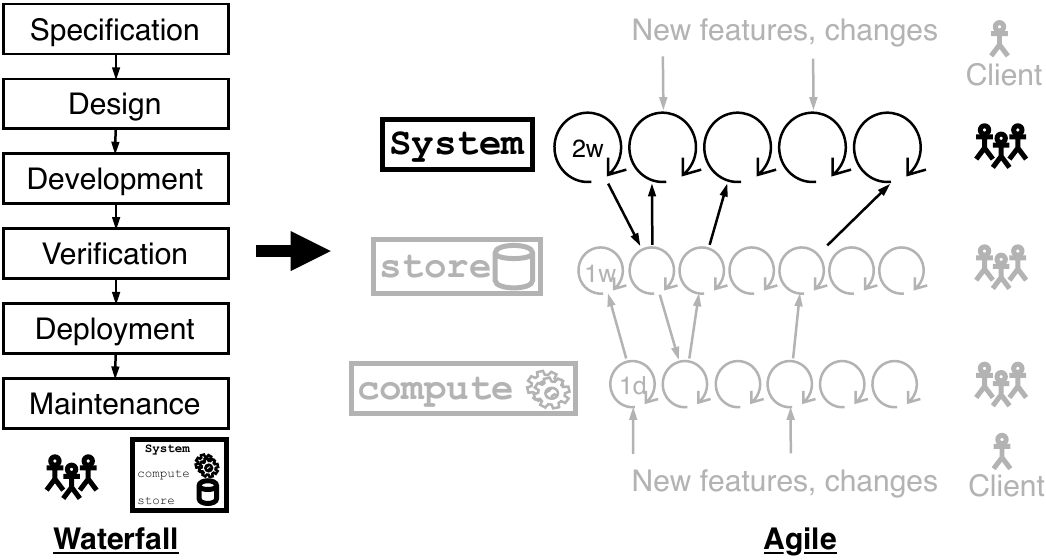}
	\caption{From Waterfall to Agile Development. \mychar{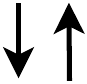}: a change from either a new version of a integrated service or from a business request, \mychar{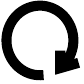}: an iteration that delivers a new version. \mychar{images/legends/team.pdf} is a development team.}
	\label{fig: agile-waterfall}    
\end{figure}

\subsection{Agile Development}
\label{sec: agile-dev}

In the 1980s and the 1990s, programmers started proposing more \emph{lightweight} models as an alternative to the sequential development process used until then that is known as waterfall development.
The traditional waterfall process consists of six phases -- specification, design, development, verification, deployment, maintenance -- executed in order. 
The new wave of agile methodologies countered this by welcoming change to come at anytime while delivering small increments of working software before all else.
For example, agile methods have ``no big design'' ethos, employ \textit{iterative} development, and focus on customer-centered engineering of functionality~\cite{beckBCCFGHHJKMMMSST2001}. 
We shorthand these methods as \textit{agile} inspired by the ``Manifesto for Agile Software Development''~\cite{beckBCCFGHHJKMMMSST2001}. Figure~\ref{fig: agile-waterfall} depicts the transition from waterfall to agile software development. 



The manifesto underlines a \textit{no big design up front} attitude \cite{beckBCCFGHHJKMMMSST2001}. 
It proposes to produce working software in short iterations (weeks, or even multiple times a day), in greater simplicity (always do the simplest thing that could possibly work and no more than what you need right now), and continuous code reviews \cite{FoxP13}. 
The phases from \texttt{design, development, verification}, and  \texttt{deployment} that in Figure~\ref{fig: agile-waterfall}, left, happen only once, are repeated every \texttt{iteration} and for all services. 
This leads to increased emphasis on modularity, incremental design, iterations in feature development that are open to \texttt{changing requirements} from the side of the business domain, and continuous deployment, i.e., releasing code as frequently as possible. 
Thus, the \texttt{business teams} on the right of Figure~\ref{fig: agile-waterfall} of every software (either representatives of the customer, or the customer) can request a change or a new feature regardless of the stage of development.

The continuous iterations, the limited scope of each produced software unit, and the modularized design enables the use of software that other teams engineer and enhances the use of service-oriented architectures; the software architecture is intertwined with the way software is produced. 
The three agile \texttt{development teams} in Figure~\ref{fig: agile-waterfall} work independently and produce the \texttt{system} and the \texttt{compute, store} services that interact with the each other through their interfaces.
This change in the teams' composition entails a lack of end-to-end \emph{control} over the system design, as other teams are responsible for the implementation behind the interfaces.
The teams' development cycles are independent.
A new increment of the \texttt{system} might be deployed every two weeks, whereas the \texttt{store} might be delivering a new version every week or every day. 
A change in the \texttt{clients} domain of each service might trigger a change in the other services that ripples as a change to the systems and services using it.
Thus, a \emph{change} might be triggered by a new version of a service integrated in a system, not only from a client's request.

Agile methods are as much about development efficiency as they are about the success of software-based businesses. 
Besides enabling customer centricity, agile methods promise to give companies control over the costs of maintenance, monitoring, and support. 
Users benefit from continuous updates instead of waiting for the release of new versions giving a big advantage in an industry where 60 percent of software costs are for maintenance, and of those costs 60 percent are for adding new functionality to legacy software~\cite{FoxP13}.

\Bt{Seda, do you think we can go without the next two paragraphs? Do you want to include them in the discussion?

Some methods make the connection between business interests and development methodologies even more evident. 
Especially development methods inspired by processes used in traditional manufacturing, e.g., Kanban, lean manufacturing were key to aligning business models, key performance indicators (KPIs) with rapid feature development driven by feedback from users or customers. 
As a result, data that could be used to measure user interactions and interest became a necessary part of optimizing features, fulfilling KPIs and pivoting companies effectively towards more promising business models. 

Finally, despite their anti-managerial gist, agile methodologies are an antecedent for iterative business experimentation, such as A/B testing, across all phases of production as a way to achieve speed and flexibility. 
In addition to enabling programming and business on demand, these methods were invaluable in addressing the inflexibility of planned development that proved to lead to grand failures and brittleness in managing uncertainty. 
Agile methodologies also promoted a much more experimental approach to software engineering, relying on user data and feedback for enabling these experiments.}

\subsection{Development and Operations (DevOps)}
\label{sec: devops}

DevOps, a term coined by combining software development and information-technology operations, has become a common practice in agile service-oriented software engineering to optimize and automate the software development process pipeline.
DevOps is characterized by a set of activities from the software development process that are being automated such as builds (from source code to a compiled version), tests (checking the conformance of new changes against a set of tests), continuous integration (integrating individual developers' changes into a single shared repository), continuous delivery (preparing new software versions that can be released), and continuous deployment (deploying new versions of the software). 
DevOps enables speedy development and deployment of new software versions with little to no engineering overhead due to the adjustable increased level of automation.
DevOps and experimental practices such as A/B testing, dark launches, or feature toggles~\cite{ibm}, and canary testing~\cite{fowler} reinforce each other.
These experimental practices make use of the multiple versions of the software that can be deployed to monitor and optimize performance indicators, i.e., clicks from users, usage pattern detection, system optimizations.
Services expose only an API to its users and thus, the many data flows and the multiple versions of the services generated through the experimental practices remain invisible from any external point of view.
The implications for privacy engineering stem from the lack of coordination and control over the data flows that may potentially lead to data leakage and linking.


\subsection{Terminology}
In the reminder of this paper, we use the following terminology. 
We use the term \textbf{service} to signify a service or a microservice, in the sense of a small computational element. 
A \textbf{monolith} is a software architecture that does not depend on externally deployed services. 
\textbf{Service architecture} is a split codebase composed of services and a service composition layer.
A \textbf{context} is the portion of the execution logic and data flows of the system to which services have access to.
\textbf{Control} expresses the level of oversight that a system designer has over the logic and data distribution of their systems, from full control on the execution logic and codebase in monolith, to partial control on both in service architectures.
\textbf{Waterfall} is a sequential software development process without loops between stages.
\textbf{Iterative} development indicates that the development process is conducted in steps that can be repeated without a constraint from which step to loop back to which other step.
At each iteration, there can be a change in the requirements towards the system in the form of either a new feature or a modification of an existing one.

\Bt{Problem for design: the API requires us to expose more information than necessary. the database stores things we don't want stored at all, such as monitoring logs}


\section{Implications for Privacy Engineering Methodologies and Standards}
\label{sec:pems_implications}

There is no doubt that the shift towards using services and agile methods facilitates the deployment of novel software-based products. 
While these approaches lower development and deployment costs, they also impose constraints on different aspects of software development that challenge the practices described in methodologies and standards that focus on engineering privacy by design. 
\Bt{Include a ref and an explanation based on Seda's work what a privacy engineering (methodology) is}

\subsection{Challenges}

We first identify three challenges that the agile and services shift in software engineering practices described in the previous Section~\ref{sec: agile} raises for methodological works and proposals for engineering privacy by design.  
A few recent studies show the impediments that software developers experience and/or express towards adopting privacy engineering methodologies in their work~\cite{HadarHATBSB18,SenarathA18,SenarathGA19,SpiekermannKL19}. 
Prominently, incompatibility between privacy engineering methodologies and usual development processes and a lack of demonstrable use are considered to be an obstacle to the adoption of privacy engineering methods~\cite{SenarathA18,SenarathGA19}.
We analyze what privacy engineering methodological works face in the current agile and services software development practices and we identify the following challenges.


\para{Challenge 1 - Dealing with services}
In service-oriented architectures, systems become a composition of \emph{services}. 
In principle, this is possibly a favorable environment for privacy. 
It may diminish the privacy risks by partitioning the data collection, processing, and storage, and facilitating the implementation of privacy mechanisms tailored to the service. 
In reality, however, services may be an obstacle to engineer privacy by design. 

First, the use of standalone services reduces development complexity by limiting the scope of what is to be done (i.e., `do one thing well'). 
However, it also increases the complexity of achieving privacy because privacy compositions are far from trivial (see Section~\ref{sec:pet_implications}).
Services may either not provide any privacy protection, or include service-specific protections~\cite{aircloak,duality}. 
Privacy engineering methods must acknowledge this difficulty and provide means and guidance to evaluate privacy of compositions of services, not only privacy of a systems under full end-to-end control of a single team.

Second, some services are provided by (untrusted) third parties or include connections to other third-party services, e.g., IP geo-location 
APIs~\cite{ipdata}, mapping~\cite{msMaps}, or face recognition~\cite{faceRecognition}.
This service-oriented philosophy is becoming also typical \textit{inside} organizations
to enable teams to work independently, only exposing APIs to each other~\cite{Yegge11}. \Bt{Remove or more to the agile Section, repetitive.}
This implicit inclusion of information flows to other entities or teams \emph{that are not under the control of the system engineer} 
raises privacy concerns that do not exist when the system is implemented as a monolith. 
Privacy engineering methodologies need to include \emph{explicit} steps to identify these 
flows, analyze them, and implement adequate protections. 
The latter step, however, may be hindered by the distributed modularized nature of the system.

Third, whether they are developed in-house or by third parties, services are typically developed by different teams. 
Services expose the APIs in their specification to other services, but hide the details of the implementation. 
This brings advantages from a development perspective, as it decouples the progress of different services and eases system updates for performance, functionality, or even security. 
From a privacy perspective, however, hiding the service' internals interferes with the overall privacy design and analysis, to the point that it may make achieving end-to-end privacy practically impossible. 
This problem is exacerbated by the fact that, because these are developed in an iterative manner, they \emph{may evolve asynchronously}. 
Such changes have to be accounted for by other services, possibly requiring new privacy solutions or changes in the existing ones. 
To address these issues, privacy engineering methodologies must incorporate steps to think about privacy requirements and solutions in pieces and increments.

\parait{Use case} Recall the privacy-preserving billing example in Figure~\ref{fig: example}. 
In the monolith solution (left), the \texttt{system} is not-trusted and protecting the users' privacy requires the use of privacy technologies \cite{BalaschRTPVG10,HoferPSK13,KerschbaumLG13,RialD11,ZhuLSHS08}. 
The introduction of services (center and right) changes the \emph{threat model}, as now one has to take into account if the \texttt{compute} and \texttt{store} services are trusted. 
In other words, the use of services \emph{introduces new information flows} with a privacy impact. 

To address this impact, the services may implement privacy technologies: \texttt{compute} may offer privacy-preserving computations and \texttt{store} may provide database encryption. 
However, the mere fact of making calls to these services can leak, for instance, the number of clients. 
This means that when services are added, one has different privacy guarantees than in the monolith scenario where only the \texttt{system} had information about the users. 
This very simple analysis highlights that once services are in place one has to conduct a new privacy analysis that must take into account the \emph{composition of flows and protection mechanisms} in the system. 

To illustrate the problems arising from \emph{independent evolution}, assume that in order to hide the number of clients, the \texttt{development team} would implement an obfuscation mechanism to hide the number of queries sent to \texttt{compute} and \texttt{store}. 
To compute the obfuscation rate, the \texttt{team} used version v1 of the services. 
After some time these services change their implementation becoming slower and handling less requests per minute (e.g., they migrate to a cheaper cloud plan). 
Even though neither the specification, nor the API of the services changes, the \texttt{system} is impacted: either it reduces the overall throughput to keep the protection level, or reduces the obfuscation rate decreasing the effectiveness of the mechanism.

\para{Challenge 2 - Minimizing trust assumptions}
The most elusive of all challenges is how to deal with the intrinsic tension between the conception of \emph{trust} in service-oriented-design and privacy-by-design paradigms. In service-oriented design software developers can, and should, use the services created by others and focus on adding value (in the form of code) where their enterprises specialize. This saves resources (time and money) and minimizes the operational risk for the enterprise. This paradigm implies that developers \emph{trust} these service to adequately handle data and protect privacy.

In privacy engineering, developers assume that the risk comes from trusting outside parties to carry out services and thus select architectures, data flows, and privacy-preserving mechanisms to minimize this risk~\cite{GursesTD15}. 
In other words, developers seek to \emph{reduce trust} in other parties to hold and protect data. 
Privacy engineering methodologies need to \emph{explicitly} acknowledge this clash between paradigms, and provide guidelines as to how to reconcile these goals. 

\parait{Use case} When introducing services to reduce costs, the enterprise who provides the \texttt{system} also externalizes the privacy risks associated to the clients' data collection and processing. Effectively, this means that the enterprise has increased its \emph{trust} in \texttt{compute} and \texttt{store}. In Figure~\ref{fig: example}, right, the \emph{trust} goes as far as to not see the flow of the \texttt{address} between the \texttt{compute} and \texttt{store} and to trust \texttt{compute} with sending the bill. Under the privacy engineering paradigm, however, a main goal is to reduce the amount of trust put on others by controlling sensitive data flows. Reconciling these two objectives raises the question of how to use services with minimal trust. A possible path is to implement more privacy technologies, but that increases the problems associated with challenge 1.

\para{Challenge 3 - Supporting development activities}
Introducing privacy-preserving technologies into software systems hinders the realization of a number of fundamental development activities, e.g., debugging and testing. In a monolithic design, these activities can be carried out prior to deployment and do not need to use production data. When iterative design and services are in place, both debugging and testing typically require the use of data collected under deployment. These data may not be available because of the privacy protections. Thus, privacy engineering must include alternative means to carry out these activities when privacy protections are in place. 

We note that these obstacles become even harder to overcome when debugging or testing concern a privacy functionality. In this case, it is likely that the data to perform the task is not available pre-deployment, e.g., setting up parameters to privately training a machine learning model on a population that has an unknown distribution, and thus cannot be used for testing. This problem is intrinsic to the very nature of the privacy solution. As such, it may be that it is not possible to enhance privacy engineering to deal with these situations.

\parait{Use case} Introducing more PETs to account for the introduction of untrusted services, also increases the difficulty to debug and maintain systems. For instance, adding obfuscating communications with the \texttt{compute} provider can hinder the debugging of billing computation; the dummy traffic will obfuscate the correct behavior of the external service for the \texttt{system} too.




\subsection{Engineering Practices in Privacy by Design Methodologies and Standards}

\begin{table*}[ht]
    \centering
    \ra{1}
    \small
    \caption{Overview of engineering privacy by design methodology contributions. Encoding: \Circle: the work does not consider or present this dimension, \RIGHTcircle: the work mentions this dimension but does not center around it, \CIRCLE: the work revolves around and analyzes in detail this dimension.}
    \label{table: epbd-methodologies-contribution}
    \begin{tabular}{@{}lccccc@{}}
        \toprule
        \textbf{} 
        & \textbf{Heuristics}
        & \textbf{Prescriptive}  
        & \textbf{\makecell{Mapping\\to PETs}}
        & \textbf{\makecell{Risks/\\threats} }
        & \textbf{\makecell{Goal-\\oriented} } \\ 
        \midrule
        
        \textbf{Academic publications}  \\ 
        
        Hoepman~\cite{Hoepman14} & \CIRCLE & \RIGHTcircle & \CIRCLE  & \Circle  & \Circle  \\ 
        
        Deng \textit{et al.}~\cite{DengWSPJ11} & \Circle & \CIRCLE & \CIRCLE & \CIRCLE & \Circle  \\ 
        
        Spiekermann and Cranor~\cite{SpiekermannC09} & \Circle & \CIRCLE & \Circle & \Circle & \Circle \\ 
                
        De Metayer~\cite{DeM16} & \Circle & \CIRCLE & \Circle & \CIRCLE & \Circle  \\ 
                
        Hansen \textit{et al.}~\cite{HansenJR15} & \Circle & \CIRCLE & \Circle & \Circle & \RIGHTcircle \\ 
        
        G\"urses \textit{et al.}~\cite{GursesTD11,GursesTD15} & \CIRCLE & \Circle  &\CIRCLE & \CIRCLE  &  \Circle \\ 
        
        Al-Momani \textit{et al.}~\cite{Al-MomaniKSKB19} & \RIGHTcircle &  \Circle & \Circle &  \Circle & \Circle  \\ 
        
        Bernsmed~\cite{Bernsmed16} & \Circle & \CIRCLE & \Circle & \Circle & \Circle  \\ 
        
        Liu \textit{et al.}~\cite{LiuYM03} & \Circle & \Circle & \Circle & \Circle & \CIRCLE   \\ 
        
        Kalloniatis \textit{et al.}~\cite{KalloniatisKG2008} & \RIGHTcircle & \CIRCLE & \CIRCLE & \Circle & \CIRCLE \\ 
        
        Kung~\cite{Kung14} & \Circle & \CIRCLE & \Circle & \Circle & \Circle \\

        Danezis \textit{et al.}~\cite{DanezisDHHMTS15} & \CIRCLE & \RIGHTcircle & \CIRCLE & \RIGHTcircle &  \Circle\\
        
        \textbf{Standards}  \\  
        
        ISO/IEC TR 27550~\cite{ISO27550} & \RIGHTcircle & \Circle & \CIRCLE & \RIGHTcircle & \CIRCLE  \\ 
        
        OASIS PMRM~\cite{PMRM13} & \Circle & \CIRCLE & \Circle & \RIGHTcircle & \Circle \\ 
        
        OASIS~\cite{CavoukianCJSDFFF14} & \Circle & \CIRCLE & \Circle & \CIRCLE & \Circle  \\ 
        
        NISTIR 8062~\cite{NISTIR8062} & \Circle & \RIGHTcircle & \Circle & \CIRCLE & \Circle  \\ 
        
        PRIPARE~\cite{NotarioCMAMAKKW15} & \Circle & \CIRCLE & \Circle & \CIRCLE & \CIRCLE \\ 

        \bottomrule
    \end{tabular}

\end{table*}

Next, we present our analysis of the privacy engineering methodologies literature with regards to what extent it covers the identified challenges.
Privacy engineering methodologies are works (frameworks, theories, methods, models, processes, technologies and tools) that are meant to aid developers to engineer privacy-preserving software. 
We selected papers whose focus is to support the engineering of privacy by design though yet their perspectives differ: certain are focused on risk assessment and threat modeling, others on requirements engineering, and few on software engineering. 
The field of privacy engineering methodologies is new and limited to only a handful of established works that we started from.
We extended the set of papers by following their references until we could not find further studies within the scope of our analysis as well as from a broader search through literature that investigates the perceptions of developers towards privacy engineering (methodologies).
We also study relevant privacy-engineering standards because of their potential impact on industry practices regarding privacy.

We excluded from our analysis non-peer-reviewed literature such as books, white papers, grey literature\footnote{Grey literature refers to literature produced outside of usual academic or commercial publication channels such as reports, working papers, etc.}. We acknowledge that some of these sources can be informative, especially about the state of practice in industry. However, they focus predominantly on niche markets of practitioners, on specific technologies, and data protection compliance, thus, are out of scope for our analysis.
We also excluded works that focus only on threat modeling and risk analysis as these are complementary to engineering activities (could be a subset) but are not sufficient as an engineering methodology. 


\para{Type of methodological contribution}
We first classify the works according to their methodological contribution (see Table~\ref{table: epbd-methodologies-contribution}). We consider the following categories:

\begin{itemize}[nosep]
\renewcommand\labelitemi{--}    
    \item \textit{Heuristics \& Thinking processes}: The methodology provides guidelines for developers regarding how to reason about privacy and privacy design.
    \item \textit{Prescriptive}: The methodology is a checklist, a taxonomy, or a reference model. The users of the methodology have to fit their case in the listed ``privacy'' cases.\Sf{unclear what it means to fit case in privacy cases}
    \item \textit{Mapping to PETs}: The methodology provides a mapping from privacy principles, best practices, activities, objectives, or strategies to one or more PETs suitable to implement them.
    \item \textit{Risk assessment/Threat modeling}: The methodology is a framework for analyzing privacy risks or for doing privacy-oriented threat modeling. 
    \item \textit{Goal-oriented}: The methodology is focused on eliciting requirements to achieve privacy goals. 
    \Bf{We no longer include this in the table but maybe it's a good idea to include it somewhere as a finding from the analysis or a discussion  CT: What do we not include????}
\end{itemize}
    
Our study highlights that methodology-oriented works \emph{rarely include guidelines} on how to engineer privacy as a software engineering practice. Most of the contributions in the literature are prescriptive, with few actionable guidelines for developers who would like to use PETs in their projects\Sf{Is there a way to make the difference more clear between prescription, guidelines and heuristics? An example?}. 
Those that provide heuristics take mainly two perspectives. Hoepman~\cite{Hoepman14} shows how to reason about privacy starting from a known software engineering concept -- design patterns -- and extending it with privacy strategies and patterns that achieve a given privacy property. G\"urses et al.~\cite{GursesTD11,GursesTD15} provide a design process in which the central point for the system-to-be-engineered is minimizing trust on entities other than the data subject. This implicitly leads to the minimization of data that needs to leave the trusted domain, providing a way for developers to reason about the boundaries of the system as well as computational methods to achieve the same outcome with less data being shared with third parties. 

The few papers that offer mappings to PETs~\cite{Hoepman14,DengWSPJ11,KalloniatisKG2008,DanezisDHHMTS15} use a \emph{top-down} approach. 
The methods start from high-level concepts (principles in the case of Hoepman's work~\cite{Hoepman14} or abstract goals in the case of Kalloniatis et al.~\cite{KalloniatisKG2008}) and go through multiple steps of transformation and refinement map to PETs to analyze and design the system and prescribe a way of working (a method or a process) that have to be followed closely irrespective of any software development practices. 
For instance, in the case study illustrating the use of the PriS method unlinkability is presented as a privacy goal, whereas ``ensure the participation
of all eligible voters'' is an organizational goal refined into multiple subgoals that are analyzed in relation to the unlinkability privacy goal; the privacy goals are also labeled as privacy patterns that map to techniques to implement them (unlinkability relates to, for instance, onion routing). 
The goals (i.e., ``ensure the participation of all eligible voters'') are disconnected from any privacy technology (i.e., onion routing) thought multiple non-trivial steps (organizational subgoals, privacy goals, privacy patterns) that have to be followed closely, and thus difficult to tailor to individual needs in custom development processes (also acknowledged in PriS~\cite{KalloniatisKG2008}).

The limitations above are transposed into standards. The standards are driven by a collaboration between a subset of the broader academic community, governmental bodies, and enterprises, and typically stress on minimizing risks for enterprises. 
In particular, we find that the threats considered in these standards (as well as the work by De and Le Metayer~\cite{DeM16}) are typically organizational, e.g., ISO/IEC TR 27550 relates to the work of Deng et al. for privacy threat modeling and lists for example, ``policy and consent noncompliance'' as a threat to be addressed~\cite{DengWSPJ11}\Cf{Betty, examples here please}. As such, they correspond to adversarial models that \emph{do not match the adversarial (threat) models} considered in academic papers on privacy technologies (such as those revised in Section~\ref{sec:pets_revision}) which focus on risks associated with information leaked by technology vis a vis adversaries (including to the system providers). Therefore, it is difficult for developers to integrate PETs when following these standards, as they fulfill different objectives than the ones set up in the threat models prescribed in standards. In fact, standards rarely mention PETs, and when they do so it is on a use-case basis with few handlers for developers to extrapolate the solutions to further scenarios.

\begin{table*}[ht]
    \centering
    \small
    \ra{1}

    \caption{Challenges and the methodologies. Encoding: \Circle: the work does not consider or present this dimension, \RIGHTcircle: the work mentions this dimension but does address the problems. \CIRCLE: the work revolves around and analyzes in detail this dimension. \grayCIRCLE: the work proposes a non-technical solution to the dimension.}
    \label{table: epbd-methodologies-challenges}

    \begin{tabular}{@{}lccccc|cc|c@{}}
        
        \toprule
        & \multicolumn{5}{c}{\textbf{\textsc{Services}}} & \multicolumn{2}{c}{\textbf{\textsc{Trust}}} & \textbf{\textsc{Developer}}  \\
        
        & \textbf{Architecture} 
        & \textbf{\makecell{Dev\\Process}}
        & \textbf{Integration}
        & \textbf{Changes}
        & \textbf{Eval}
        & \textbf{\makecell{Context\\of use}} 
        & \textbf{\makecell{Minimize\\Data}}
        & \textbf{\textsc{Activities}}\\ 
        \midrule
        
        \textbf{Academic papers}\\
        Hoepman \cite{Hoepman14} & monolith & \makecell{mentions \\iterative} & \Circle & \Circle & \Circle & \Circle & \RIGHTcircle & \Circle  \\
        
        Deng \textit{et al.} \cite{DengWSPJ11} & monolith &  waterfall & \Circle  &  \Circle  & \Circle & \Circle & \Circle & \Circle  \\ 
        
        \makecell[l]{Spiekermann\\and Cranor \cite{SpiekermannC09} } & hybrid & waterfall &  \Circle & \Circle & \Circle & \Circle \grayCIRCLE & \RIGHTcircle & \Circle     \\ 
                
        PRIAM \cite{DeM16} & hybrid & waterfall &  \Circle &  \Circle  & \Circle & \Circle & \RIGHTcircle & \Circle  \\
                
        Hansen \textit{et al.} \cite{HansenJR15}  &  hybrid& waterfall &  \Circle &  \Circle  & \Circle & \Circle & \RIGHTcircle & \Circle  \\ 
        
        G\"urses \textit{et al.} \cite{GursesTD11,GursesTD15} &  monolith &  waterfall &  \Circle & \Circle & \Circle & \Circle & \CIRCLE & \Circle  \\ 
        
        Al-Momani \textit{et al.} \cite{Al-MomaniKSKB19} & monolith & v-process\footnotemark   &  \RIGHTcircle &  \Circle & \Circle & \Circle & \Circle  & \Circle  \\ 
        
        Bernsmed \cite{Bernsmed16} & monolith & waterfall & \Circle & \Circle & \Circle & \Circle & \Circle & \Circle  \\ 
        
        Liu \textit{et al.}  \cite{LiuYM03} & monolith & waterfall &  \Circle &  \Circle & \Circle & \Circle  \grayCIRCLE  & \Circle & \Circle  \\ 
        
        Kalloniatis \textit{et al.} \cite{KalloniatisKG2008} &  monolith & waterfall &  \Circle &  \Circle & \Circle &  \Circle & \Circle \grayCIRCLE &  \Circle  \\ 
        
        Kung \cite{Kung14} & monolith & waterfall & \Circle & \RIGHTcircle & \Circle & \Circle & \Circle & \RIGHTcircle  \\

        Danezis \textit{et al.} \cite{DanezisDHHMTS15} & hybrid & \makecell{mentions \\iterative} & \RIGHTcircle & \Circle & \RIGHTcircle & \Circle  \grayCIRCLE  & \RIGHTcircle & \Circle  \\
        
        \textbf{Standards}  \\ 
        
        ISO/IEC TR 27550 \cite{ISO27550}  & services & \makecell{mentions\\agile} & \Circle  \grayCIRCLE  &  \Circle  \grayCIRCLE & \Circle & \Circle  \grayCIRCLE  & \Circle & \Circle   \\ 
        
        OASIS PMRM \cite{PMRM13} & services  & waterfall & \Circle &  \Circle & \Circle & \Circle \grayCIRCLE & \Circle & \Circle  \\ 
        
        OASIS \cite{CavoukianCJSDFFF14}  & services & \makecell{mentions\\agile} & \Circle & \Circle & \Circle& \Circle  \grayCIRCLE & \RIGHTcircle & \Circle  \\ 
        
        NISTIR 8062 \cite{NISTIR8062}  & monolith & waterfall & \Circle & \Circle & \Circle & \Circle  \grayCIRCLE & \Circle & \Circle \\ 

        PRIPARE \cite{NotarioCMAMAKKW15}  & monolith & agile & \Circle & \Circle & \RIGHTcircle & \Circle  \grayCIRCLE & \Circle  & \Circle  \\ 

        \bottomrule
    
    \end{tabular}

\end{table*}

\para{Addressing the challenges} Next, we evaluate the extent to which these methodologies address the challenges that the introduction of agile and services pose. We evaluate the works according to the following criteria, which reflect the challenges we identified in the beginning of the section (Table~\ref{table: epbd-methodologies-challenges} summarizes our analysis).  

\begin{itemize}[nosep]
\item\textit{Addressing services.} We consider five dimensions that methodologies need to address. 
(1) \textit{Architecture}, what architecture does the methodology assume? a monolith (a single codebase with no modularity and no coordination with external entities), service-oriented (a traditional service-oriented architecture) or a hybrid (the methodology mentions services but the architecture is not assumed to be \emph{service-oriented}). 
(2) \textit{Development process}, does the methodology assume (implicitly or explicitly) a certain type of development process for the engineering of the software? 
(3) \textit{Integration}, does the methodology propose guidelines for integrating services together? 
(4) \textit{Changes and new features}, does the methodology consider and/or provide a way for engineers to reason about changes in the system? 
(5) \textit{Evaluation}, does the methodology address the evaluation step, and in particular does it acknowledge the difficulty of carrying out a privacy analysis when PETs and threat models of different services need to be combined?

\item\textit{Addressing the shift in trust assumptions} is reflected in two dimensions. 
(1) \textit{Context of use}, the methodology inspects and provides ways to reason about the contextual dependencies between services: which part of the system is responsible for what part of the execution logic and which data flows. In this dimension, we also distinguish between the context that service-oriented systems exist in (technical settings for the engineering systems with services) and the non-technical organizational context of use of information systems. Methodologies that focus on the organizational context might be more inline with contextual integrity~\cite{Nissenbaum18}. 
(2) \textit{Data minimization}, the methodology addresses data minimization as a prime system design principle by looking at ways how to fulfill functionality in a way that minimizes the need to trust other parties or a single party, with the protection of data~\cite{GursesTD15}. 
\Cf{These two dimensions need to be better written/justified.Please check changes}

\item\textit{Addressing development activities.} We consider whether the methodology acknowledges that privacy engineering has to accommodate development activities (such as debugging, testing, continuous integration) beyond the pure design decisions and implementation of protections in a system. Or, does the methodology provide support for developers to perform development activities while preserving the privacy properties of the system?

\end{itemize}

\parait{Academic publications} Regarding services, we observe that 66\% of the academic publications assume that the system is a monolith controlled by one entity, and the remaining works only mention services but do not really consider modularity or flows between entities. For instance, Spiekermann and Cranor~\cite{SpiekermannC09} discuss services, from the point of view of the user: the system provides a service to the user. Their work assumes that the service provider have full control over the functionality and implementation of the service. Similarly, other works that introduce the service notion~\cite{Bernsmed16,DeM16,HansenJR15} focus on the user perspective, which is far from the principles underlying service-oriented architectures, e.g., modularity or reliance on third party service providers.

We find a similar breakdown in terms of development process, 75\% implicitly or explicitly assume waterfall methodology in which design is made up front and never changes, and the others do mention iterative development although they do not explicitly address the challenges that arise from that change. In fact, only the work by Danezis et al.~\cite{DanezisDHHMTS15} acknowledges explicitly that how to deal with changes in the development process is an open question. \Sf{Rephrase this last sentence to make it more clear ->}Interestingly, there seems not to be a clear pattern that matches architecture and development process, which highlights the disconnection with respect to current software development practices where those go hand in hand.

The treatment of integration, changes, and evaluation is scarce. Integration and changes are mostly ignored, even by those works that acknowledge the existence of services or iterative methodologies. The most sophisticated treatment of integration is the requirement on privacy technologies to be ``integrable'' and ``compatible'' by Al-Momani et al.~\cite{Al-MomaniKSKB19}.\Sf{Unclear sentence for people who haven't read the paper ->}\Bf{better?}
However, integrable in this paper is not related to actual integration of services but to whether the PET offers the functionality and performance needed in the system. The example that illustrates the notion of a non-integrable technology is a computationally-heavy homomorphic scheme and how it cannot be integrated in a time-critical application. This does not show impossibility for integration but inappropriateness, hence, even this work does not review the aspects of integration that are intrinsic to service-oriented software. The last aspect, evaluation, is mostly equated to a Privacy Impact Assessment~\cite{Clarke09}, in which designers evaluate the impact of a privacy violation \textit{but they do not quantify the likelihood of such violation given the system specification}.
Evaluation, as understood in academic papers on privacy technologies, is only explicitly mentioned by Danezis et al.~\cite{DanezisDHHMTS15}, where they point out that privacy evaluation must be addressed, but they do not explain how nor mention the difficulty behind composition of service-specific privacy protections.

Our analysis indicates that development activities such as testing, debugging, continuous integration are ignored across the board. Kung~\cite{Kung14} mentions tangentially that there are tactics on the software architecture level that can be employed to achieve ``modifiability'' and ``testability'' among other properties. As he proposes a software architectural solution for privacy engineering, we could interpret that the existing tactics are compatible with the privacy tactics posed in the work. Yet, there is no stated relation between and analysis of the combination of the two.

\footnotetext{Considered an extension of waterfall. The v-process bends the activities upwards after coding to show the relationship between steps.}

\Sf{I think I know what you want to say in this paragraph, but needs to be more clear} \Bf{did I do better?} Finally, regarding minimizing the trust put in any single party, a few works do explicitly mention the system's \emph{context}. However, this context is the settings of the organization not the notion of context of a software system. For instance, Liu et al.~\cite{LiuYM03} model the organizational context through actors and their goals. Data minimization, on the other hand, is often mentioned~\cite{GursesTD15,GursesTD11,Hoepman14,SpiekermannC09,HansenJR15}, but it is central only to the methodology proposed by G\"urses et al.~\cite{GursesTD11,GursesTD15}. It is, however, only their later work that presents and expands on the concept of trust minimization as pivotal in privacy-preserving systems design~\cite{GursesTD15}.

\parait{Standards} We observe a different trend on standards than on academic works. The majority of standards acknowledges the existence of services, in the service-oriented architectures sense, and some of them do consider integration and changes. In particular, ISO/IEC TR 27550~\cite{ISO27550} provides a detailed account how to manage the relationship with subsystem providers and that the expectations and contractual obligations between the users and the main system provider should extend to the subsystems. This recommendation is a clear sign of service-oriented architectures being taken under consideration, yet, the focus remains on contractual issues such as defining liability and responsibility. The standard stops there and does not tackle the complexities arising from services and privacy integration, and hence, the challenges we raise in this section.

Notably, the standards we reviewed center around the context of use of an IT system and place the analysis and design of this context in the heart of the standards. However, standards are primarily concerned with the organizational aspect of IT and how IT is integrated and standardized within this organizational context. 
This organizational focus limits the scope and purpose of standards to high-level general recommendations and instructions on conducting privacy engineering within the organizational context.
The implications of the software development process together with the services architecture could be interpreted as too specific to be included in standards; after all standards are industry-wide documents which should be applicable regardless of the type of the software development process employed or the software architecture type.
Yet, in omitting these specificities, the standards also fail to address the challenges that the predominant software development process (agile) and software architecture (service-oriented) pose to the actual integration and engineering of privacy in systems.
\Sf{this paragraph could be much sharper and have a great final punch line}\Bt{help, if you can}


\section{Implications for Privacy Enhancing \\Technologies}\label{sec:pet_implications}

The shift to iterative software engineering and the reliance on services pose challenges not only to the privacy engineering methodologies but also to the design of privacy enhancing technologies. 

\subsection{Challenges}

We now describe challenges and discuss whether their origin 
is the use of agile or the use of services.

\para{Challenge 1 - Changing requirements}
In an \textit{agile software development}, the specification of the system is never assumed to be final. 
At every iteration there can be a change in the target functionality, the amount or type of data that is processed, or in the services used in the system. 
Such changes effectively result in a new system specification. 
PETs designers will have to account in their designs that a change can, and inevitably, will occur during the development of the overall system. 
The system specification change might result in a change of the threat model, the requested privacy protection, the requested functionality (new, removed, reduced or enhanced), or the way the privacy-preserving functionality is carried out.

\parait{Use case} Recall the privacy-preserving billing example in Figure~\ref{fig: example}. 
Assume that the customer buying the system requests a change in functionality: add a new feature to the bill computation (e.g., a time slot, 
a per-client premium, or new data collection). 
To implement these changes, one must change the  encrypted information, as well 
as any zero-knowledge proof used to prove that the encryption is correct.
These changes, however, are far from trivial and may require a new design from scratch.

\para{Challenge 2 - Agnostic privacy protection without control}
Partitioning the execution logic and data flows of a monolithic system into services means that the PETs that are integrated in different parts of the service-oriented system will have a limited overview of the context of the end-to-end system, thus, the PETs designers do not have full control over system. 
The next challenge for the PETs is to offer privacy protection to be integrated with the different services without having control over, or even knowing, the rest of the system (i.e., services used, execution logic, data flows).
Depending on the end-to-end system, services may gather different types of data, for different purposes, and be considered in different adversarial roles. 
Therefore, the PETs that will have to provide protection independently from the end-to-end system: the PETs can guarantee only minimal limited protection.
This problem is exacerbated by the widespread use of certain services, e.g., OAuth-based login services~\cite{facebook,google}
or cloud-based Customer Relationship Management software~\cite{Salesforce,Nimble}.
By virtue of being used in more than one context, these services have access to unexpected data flows generating privacy, in particular linkability, issues that are hard to foresee for both the service and the end-to-end system designers who have only partial visibility of the data flows.
The PETs will have to be system-agnostic. 
Meaning, the PETs will have to ignore the specific data that are being gathered, the interaction between services, the architecture of the application and its specificities, the compositions of services and PETs.
PETs are highly contextual because they provide concrete privacy guarantees under a specific settings (thread model and deployment setup); deploying PETs in service-oriented systems would require the PETs to abstract these specificities and provide guarantees in an agnostic to the end-to-end system way.

\parait{Use case} For example, in Figure~\ref{fig: example}, center and right, the \texttt{compute} can be coupled with a PET that provides a guarantee that the \texttt{premium} is never seen in plain text outside of the service. The moment the \texttt{compute} sends the \texttt{premium} outside of its boundary, the privacy protection is outside of the control of \texttt{compute}. Now consider the development team behind the \texttt{compute} service (Figure~\ref{fig: example}, center). 
While they are designing and implementing, they cannot be sure about (i) which will be the requirements
of the system (Challenge 1) and (ii) under which environment, i.e., with what other services and under what configuration,
will the module be deployed. Therefore, the team cannot predict what other flows there will be under deployment
and implement the adequate safeguards to ensure that the privacy guarantees provided by the privacy-preserving
computation are preserved regardless of the environment.


For a PET to be integrated in a service, it usually means it is weaved into the code of the service. In other words, the PET and the service are not separated but one (for example, an encrypted database).
To allow for services to integrate PETs more easily (the encryption), the PETs have to provide an interface: the PETs designers need to abstract the details of the implementation and generalize the cases in which the PET will be used.
The questions to address are many: what a meaningful abstraction is, what the boundaries of the PET and the service are, what functionality that the PET assumes, how much to expose from its internals, how to be interoperable and backward-compatible.

\parait{Use case} Now consider the development team behind \texttt{compute} (Figure~\ref{fig: example}, center and right). 
In the first case, (Figure~\ref{fig: example}, center), the \texttt{compute} service receives as an input \texttt{m}$_i$, a measurement, and returns a \texttt{premium}. 
Whereas in the second case, (Figure~\ref{fig: example}, right), the service assumes more functionality and actively asks for data from the system: \texttt{compute} requests the measurement \texttt{m}$_i$ after receiving an \texttt{address}, and sends a \texttt{bill}, never exposing the premium to another service. 
These two cases differ drastically in their threat models and context that the services have access to. 
If \texttt{compute} has no privacy properties to guarantee, it can expose two different entry points for its users to decide which of the two cases they prefer. 
However, in the case of a privacy-preserving computation, the two scenarios might contradict each other's threat model (in the first case, the \texttt{address} and the \texttt{premium} are associated by the \texttt{system}, in the second, the \texttt{system} never receives either pieces of data).

\para{Challenge 3 - Support composition}
Building service-oriented systems would mean that more than one PET are likely to co-exist.
However, PETs are typically designed and evaluated to provide a given privacy protection in \emph{one} specific system setup.
A PET designer would have to anticipate the need for composition and evaluation of compositions of PETs with other PETs and traditional services.
This means that the designers would have to make use of privacy definitions and adversarial models that can be composed and that provide grounds for evaluation of the privacy guarantees of the end-to-end system.

\parait{Use case} Even if it is the overall system designer responsibility to make sure that when integrating services they do not reduce each others' protection, this can only be done if the PETs implemented in those services can be composed. For instance, assume that the \texttt{store} service permits retrieving records under oblivious transfer. 
If the oblivious transfer and the homomorphic billing do not share a threat model, studying the composition is hard. 
An open question is whether it is possible to generalize and to express threat models and consequently privacy and security definitions that can be composed and evaluated.

\para{Challenge 4 - Developers activities for and with PETs}
Besides core design and evaluation tasks discussed in the previous challenges, system development and deployment entails other activities to ensure, among others, correct operation (e.g., functionality testing) or adequate performance (e.g, collection of usage statistics).
These practices introduce new requirements for PETs. For instance, PETs that are integrated in a service may be obfuscating the data flows, encrypting the data, introducing noise, etc. to provide privacy protection to the service.
These techniques will transform the PETs into a near black-box for the developer who integrates it with their service as the behavior of the service will be altered by the PET and the developer will not be able to inspect the inner workings to detect malfunctions or incorrect use.
The PETs designers need to account for reproducibility, traceability, rigorous behavior specification to enable the developers to integrate PETs without disrupting usual developers activities.
PET designers will need to come up with creative novel and certainly non-trivial ways to ensure testing of functionality and reproducibility.

Other development activities are associated with DevOps practices. 
An example are the uses of canary testing and dark launches
in which different versions of the software co-exist for functionality or performance testing.
These technologies change flows and introduce unforeseen flows which are never considered by PETs designers.
This hinders PETs deployment as they can not support modern deployment needs.

\parait{Use case} In our privacy-preserving billing running example, for instance, the \texttt{system} team needs to know that the \texttt{compute} service performs up to expectations: that the \texttt{premium} is computed correctly. Yet, if the \texttt{premium} integrates a PET that encrypts the value, then the PET designer will have to provide additional mechanisms to prove the operation was correct. This can be achieved with, for example, a zero-knowledge proof but this is in no way trivial or straightforward to implement (by the PETs designers) or integrate (by the \texttt{compute} developers).

\subsection{Engineering Practices in Privacy Technologies Research}
\label{sec:pets_revision}

In this section, we study whether researchers working on PETs
and privacy-preserving systems address the challenges in the
previous sections and if they do so, how.

\para{Paper selection} We conducted a systematic literature review of privacy-related papers. 
We selected these papers by revising the programs of three top
conferences: IEEE S\&P, USENIX Security, NDSS
between 2015 and 2019. We identified 118 papers (33 from
IEEE S\&P, 41 from USENIX Security, and 44 from NDSS). 
The selection comprises papers presented in 
privacy-oriented sessions, as well as other papers that
advance the state of the art on privacy. We did not consider
papers that propose, improve, or analyze cryptographic primitives or
their implementation unless the authors presented them 
in the context of a clear privacy application. We also
excluded from our study papers focusing on privacy policies (see Section~\ref{sec:related}).
We performed a cursory review of privacy-related papers from ACM CCS and PETS.
Our sample confirmed that the papers published in those two venues fare similarly to the other papers that we study against the challenges.
Therefore, we stopped our papers sampling because our analysis reached a saturation point: 
(1) increasing the volume of papers included in the analysis did not bring further insights about the treatment of agile and services in the PETs-oriented literature; and (2) the percentages presented in the remainder of the section remained similar even if the absolute numbers increase. 

\para{Nature of the contribution}
We study the type of contribution these papers make as 
building blocks that can be used for privacy engineering.
We consider five categories:
\begin{enumerate}[nosep]
	\item \textit{End-to-end systems}: these are papers that propose a complete
	system to solve a given problem in a privacy-preserving way
	(e.g., messaging~\cite{Corrigan-GibbsB15}, ride-hailing services~\cite{PhamDETHH17}, 
	or credit networks~\cite{MalavoltaMKM17}). 
	These papers consider the complete `development cycle', eliciting requirements,
	designing an end-to-end solution, implementing it, and performing an evaluation 
	of its privacy properties and the overhead these impose. 
	\item \textit{Components}: these are papers that propose a component that
	enhances privacy for one particular functionality (e.g., a measuring unit~\cite{JansenJGED18,ManiS17},
	traffic obfuscator~\cite{WangG17}, or search~\cite{MishraPCCP18}). 
	These components can be integrated in systems to enhance privacy.
	\item \textit{Protocol}: these are papers that propose a protocol or an algorithm
	that enhances privacy, but cannot do this all on its own~\cite{WangBLJ17,AngelCLS18}. 
	\item \textit{Evaluations/Analysis/Attacks}: these are papers that do
	not provide new privacy-preserving functionalities, but conduct
	evaluations to test the privacy of existing systems, components or 
	protocols~\cite{PyrgelisTC18,SchusterST17,VenkatadriALMGL18}.
\end{enumerate} 

We assign papers to one or more categories (e.g., papers that propose end-to-end systems
may also introduce reusable components). We end up with 18 end-to-end systems, 40 components, 
20 protocols, \textit{77 evaluations and attacks}. 
Grouping these, we obtain 78 papers focused on design (end-to-end, components, and protocols) and 77 
evaluation-oriented. In other words, 50\% of the privacy work at top privacy conferences
focuses on attacks and evaluations of systems. While this helps increasing security, 
privacy, and transparency on academic and deployed systems, it brings little value
to the privacy engineering discipline.

\para{Consideration of engineering aspects}
We look at engineering aspects that are key for the integration of privacy technologies in 
contemporary software development pipelines.
We consider the following aspects:
\begin{itemize}[nosep]
	\item \textit{Systematization}: The paper reviews a (part of) the literature 
	and provides an organization/comparison/systematization 
	(e.g., a taxonomy, a hierarchy, a mapping to the attack/design space)
	that would help engineers to identify solutions~\cite{JiLMHB15,RazaghpanahNVSA18,SunEVLRCM15}. 
	\item \textit{Generalization}: The paper makes an effort to generalize the 
	use of the proposed techniques beyond one or two use cases~\cite{DyerCS15,LachariteMP18,MelisDC16}.
	\item \textit{Sets best practices}: The paper indicates steps that should be 
	taken to best design/evaluate a system~\cite{BartonWMI18,SonKS16,VenkatadriALMGL18}.
	\item \textit{Frameworks}: The paper proposes a guidelines that
	can be followed when designing, implementing, or evaluating a system to 
	reproduce the steps taken by the authors should the environment,
	adversarial model, or other aspects, change~\cite{JiLMHB15,LachariteMP18,MaticTC17}.
	\Bt{challenges 1,2,3,4???}
	\item \textit{Reference code/artifact}: The authors have made reference code available.
	(We do not test whether the code runs or whether the results are reproducible).
	\Bt{challenges 4?}
	\item \textit{Considers context and integration}: The paper speaks about the environment 
	where the component/protocol will be deployed~\cite{Delignat-Lavaud16,MelaraBBFF15}.
	\Bt{challenges 3}
	\item \textit{Considers changes}: The paper considers that the requirements, 
	components, services, or environment may change over time and discusses how
	these changes will affect the design and its properties~\cite{BostPTG15,HarkousFLSSA18}.
	\Bt{challenges 1, 3}
\end{itemize} 

In terms of laying out the research space around privacy technologies, we find that
only 15 (12\%) papers provide some sort of systematization, and 41 (21\%) elaborate on 
generalization beyond demonstrating effectiveness for the proposed methods in one or
two concrete use cases.
We obtain a similar result when looking for descriptions of the process
followed by the authors. Only 15 provide a framework to guide engineers 
to tackle similar problems, and only 32 (20\%) establish best practices
that can be followed by engineers to provide strong privacy protection
in their systems. 

Among the 118 papers, 56 (47\%) provide reference code or an artifact associated
with the research. However, only 23 of these 56 correspond to protocols and
components. The remaining 33 correspond to an attack or an evaluation.

If we look at agile and service-oriented aspects such as consideration of
context and changes, the results are even more discouraging. Only 23 (19\%)
papers explicitly mention the environment in which technologies could
be deployed; and only 13 (11\%) papers consider that this environment
may change over time (and these sets have two papers in common).

When we consider papers that propose components, which by definition
are to be integrated in larger systems and thus should be more cognizant of
environmental conditions, the results improve but are still far from ideal.
Out of 40 components 10 (25\%) explicitly mention the context where
they would be deployed, but only 3 (7.5\%) are aware that this context
may evolve over time. 
We found only six 
evaluation-oriented papers that consider context and only
one explicitly address changes.
This means that while these papers represent 50\% of the academic output, 
they focus on snapshots of systems and their broader applicability in
software engineering is in question.

The most discouraging result is that 44 (37\%) of the papers we survey \textit{do not mention
any of the considered engineering factors}. This includes 18 papers
that do provide reference code but do not discuss how this code may be integrated
in software engineering practices.



\Sf{ we removed two sections here. They are commented out. Check commented out text to see we covered everything.}

\Sf{ there is a paper \cite{RubinsteinG13} that has already discussed the unsuitable nature of the PbD principles/methodologies w.r.t. these being too abstract (quote: aspirational rather than practical or operational) as well as not being validated, meaning: there are no studies on the usefulness of PbD that measure the effect that such a design methodology would have.}


\section{Scope} 
\label{sec:related}

The purpose of this paper is to formulate and analyze systematically the challenges that software engineering practices pose to privacy engineering, hence we focus on proposals for privacy technology design and privacy engineering. Regarding privacy technologies (PETs) we covered theoretical and practice-oriented research that engages in architecture, protocol, or component design that achieve desired functionalities in a privacy-preserving manner. PETs are technologies that protect the opacity of the individual when using information services. PETs consider entities that collect and process data to be a single point of failure for privacy and propose designs intended to reduce the dependency on this centralized entities. To achieve this, privacy designs propose \emph{server-side solutions}, i.e., deployed by the service provider such as homomorphic encryption; \emph{client-side solutions}, i.e., deployed by the user to protect their interactions with service providers such as anonymous communications; and \emph{collaborative solutions} in which data and computation are distributed among the users such as peer-to-peer file sharing~\cite{DiazTG13}. All these designs aspire to minimize the privacy risks associated with data collection and processing by minimizing collection, replication, linkability, and centralization of user data when building systems. We considered privacy technology design approaches that have data and risk minimization core to their proposal.
Regarding privacy engineering, we covered research and practical guidance on methods, techniques, and tools that aid in engineering systems that integrate privacy technologies into systems to enhance privacy protection~\cite{GursesDA2016}. 
We focused on the technical feasibility of the proposed designs and on the actionability of the methodologies in current day software development environments and not on the appropriateness of these approaches for current day privacy needs. 


There are other proposals for privacy technology design and engineering that we left out of our study of this paper. First, we did not consider information accountability tools developed to support data protection compliance. In addition to limiting data collection, data protection laws are about the transparency and accountability of organizations that collect and process personal data. The EU General Data Protection Regulation~\cite{gdpr}, for example, lists a number of principles intended to make organizations that collect and process personal data accountable by making these activities transparent towards data subjects, regulators, and the greater public. These requirements can translate to data lifecycle management tools that enable the formalization of privacy policies~\cite{BreauxHR14} or ensure purpose-based access control~\cite{PetkovicPZ11}. Other works propose ways to improve transparency requirements like data subject access requests and data portability. These proposals typically assume a \emph{trusted service provider} and so far lack normative assertions with respect to data minimization in the original design of the service. These works are valuable for the the fulfillment of broader data protection requirements, but are complementary to the privacy engineering methods and privacy designs we considered.
 

Second, we did not cover research on the \textit{usability for users} of privacy functionality, anywhere from the accessibility of privacy policies~\cite{SadehABC14} to expressiveness of privacy settings~\cite{LinLS14}. This research is extremely valuable for the success and evaluation of privacy technology designs and merits expertise and analysis beyond the limits of this paper. Third, we do not cover \textit{usability of security and privacy for developers}, e.g., how they can use privacy preserving functionality, or how teams can be organized to manage implementation and deployment of security and privacy functionality~\cite{AcarFM16}. This line of research is deeply engaged in current day software engineering practices, yet it focuses on another level of abstraction: they look at developers' daily practices and usability of privacy and security functionality for developers. This research is thus orthogonal to our focus on the specification, design and implementation of privacy functionality, considering human factors as one dimension in our concerns.

\section{Discussion}
\label{sec:discussion}

Proposals for privacy design and engineering fall short on many aspects that are essential to current software engineering practices. 
In the following, we discuss the underlying gaps, and list challenges that need to be addressed to reconcile privacy and software engineering practices. We note that although these challenges have strong technical components, they are potentially impossible to address using technical means only, requiring academics, practitioners, regulators as well as the software industry to revisit current practices.

\subsection{Mind the Gaps}
\para{The Computer Science and Engineering Schism}
Computer science researchers propose designs and methodologies that often abstract away engineering aspects with the ideal of presenting results that are valid independent of the material conditions of production or engineering. Abstraction is one path to identifying designs and methodologies that are more generalizable. Moreover, researchers like to explore novel ways of doing things, rather than constraining themselves to current industry practices. For example, most PETs aspire for solutions that avoid the dominant political economy in current software industries which relies on extractive models of data collection at the cost of privacy. 

The desire to be compatible with current software engineering practices can be difficult and at times counterproductive. Fundamental research may require a different speed than the industry, simply making it difficult to keep up with changes in the software industry. Companies may keep their engineering practices and privacy challenges secret. Finally, developers and engineers may feel better equipped to figure out how to best implement proposed designs in their working environments. This makes the ideal cut off in specificity of a privacy engineering methodology an interesting matter to study.

All these justifications aside, we show that the gap between privacy engineering solutions and actual software engineering practice also exists because research proposals assume monolithic designs and waterfall methods as the mode of software production. The abstractions used in research works hold certain assumptions about how software is produced~\cite{Birnhack12}. The papers included in our study do not address challenges associated with the very nature of how software looks like and is developed. Applying solutions that are disconnected from the agile and service-oriented nature of software, leaves too wide a gap for software developers to close. Privacy engineering in the context of services, requirements and implementation evolution, changing trust models and environments, supporting developer activities, and multiple contexts of use is far from trivial and still not a solely engineering task that developers are able to perform without fundamental research in the area. These gaps raise challenges to the operationalization of the proposed designs and methods, but also open up new research questions.

\para{Silos across sub-disciplines} Some of these implicit assumptions could be surfaced through greater cooperation across sub-disciplines in computer science. In particular, researchers working on privacy designs and engineering may benefit from collaborations with researchers in software engineering.
However, currently, these communities rarely overlap and have different conceptions of privacy and the solution space. 
For example, in the software engineering field, researchers have studied engineers' perception of privacy (cf.~\cite{HadarHATBSB18,SpiekermannC09}) and requirements engineering for privacy (cf.~\cite{Ayala-RiveraP18,ShethKM14}. Others have focused on information accountability, e.g., on translating legal documents to requirements (cf.~\cite{BreauxHR14}) and the effect of regulations on systems, discovering that high-level compliance is hard to translate into technical requirements~\cite{ShahBSWC19,ShastriWC19}.
The software engineering research field holds certain implicit beliefs -- the same way privacy engineering holds waterfall and monolithic architectures -- about privacy and about how systems are engineered. Privacy in these works usually remains on the level of data protection laws and challenges to comply with the law in technical designs. Moreover, as the mode of agile and service-orientated architectures is adopted throughout software engineering field, the modus operandi is that this is the way software is engineered; hence, software engineering research holds agile and services as implicit beliefs and does not consider the implications for privacy engineering not operating under the same assumptions.
A convergence between these fields may enrich methods and help develop privacy engineering methodologies and designs fit for modern software engineering .

\para{Standards amidst many currents}
Current efforts in standardization suffer from the same limitations as academic works. For example, the stages in the life cycle are described in standards in terms of a purpose and outcomes, e.g. \textit{``Verification and validation documentation''} in the development phase, or \textit{``Quality assured systems-of-interest accepted by the acquirer''} in the production phase. Both examples implicitly assume that there is a testing activity going on. However, the description of this activity is hard to map back to practice. For instance, development verification documentation may refer to the test themselves or their results; and quality-assurance provides no guidance regarding when or how this test should be performed. In contrast, current practices perform continuous testing \emph{both} under development and production. Furthermore, ideally, standardization should come after a technology is mature enough~\cite{Waldo06}. Given the gaps between computer science and actual engineering practice, current efforts may standardize misaligned approaches, leading to a missed opportunity for an industry wide change.

\subsection{The Elephant in the Room: The Agile Turn as a Problem}

The challenges that services and agile methods poses are difficult and beg the question whether something has to change in the way we produce software in order for privacy engineering to be feasible in the first place. There is little to excuse for privacy researchers' lack of attention to software engineering practice, but any further efforts need to also consider how current software industry practices have to change.

Most importantly, it is possible to argue that service architectures and agile development methodologies have become the source of many privacy problems. Privacy research generally, and web measurement more specifically, has typically looked at advertising as the main driver for privacy intrusive development practices~\cite{AcarFM16,EnglehardEZRNt14,Vallina16,Zuboff19}.
From our vantage point, the irruption of software in the advertisement industry, and in particular the tools to enable massive tracking can be depicted as the first successful scalable product of service architectures. 

Potentially, the advertisement tracking ecosystem can be considered an example of the greater developments in software production. The capability to continuously tap on user interactions as feedback, combined with agile development methods, enable unprecedented optimization of systems~\cite{KulynychOTG20}. This feedback is key for customer and user-centric development, continuous delivery, and rapid integration of third party services. These practices are not just steps in development methodologies that can be changed, but are baked into development environments, tools and cloud service offerings. 

The distributed architecture underlying services, combined with data-centric development methods and environments, ends up multiplying privacy risks. In privacy literature, distributed systems are posited as a way to decentralize data, computation and therewith power asymmetries. However, in service architectures user data captured by one service is often shared with all the others. Companies small and large inherit these practices every time they plug into service ecosystems. These ecosystems are economically attractive for lean startups that want to test minimal viable products and for industry players that want to minimize risk while pivoting business models. Data-centric development practices using current service architectures means that privacy risks are abound regardless of the size of the company or the application domain. 

Service ecosystems and cloud infrastructures disadvantage organizations that want to use privacy designs and methodologies. Privacy design and methodologies, by design, limit the way in which the advantages of service ecosystems can be leveraged. For example, let us assume a company is committed to applying privacy engineering, embedding PETs in their design, and pursuing a different business model. Their developer teams will have to rely on non-standard software practices. This will require software developers who are willing to work without common toolsets and methodologies. Their users may have to tolerate systems that may lack some of the ``seamless'' experiences of data-centric designs. Investors will have to be dedicated to support such efforts. Setting up such a company, or bootstrapping privacy preserving development environments may turn out to be an uphill battle.

The diametrically opposed use of the same terminology in the software industry and privacy approaches surfaces the tensions that are at play here. Ease of service \textit{composition} increases complexity in privacy \textit{composition}; the extension of \textit{trust} to all third party service providers transforms services from `single points of failure' to `many points of failure', and externalizing business \textit{risk} and \textit{costs} comes at the \textit{cost} of increased privacy \textit{risks} and engineering \textit{costs}. 

Finally, even if we could consolidate terminology, practices and development environments, service architectures have gone hand in hand with the centralization of data and computation in the hands of a few dominant players like Google, Microsoft, Facebook and Amazon, making privacy engineering in this setting potentially ineffective. Some of these services have deployed privacy preserving methods, e.g., differential privacy. However, having access to immense amounts of data and creating computational dependencies can yield tremendous power, \textit{even} when the data are processed by these entities in a privacy-preserving manner~\cite{Rogaway15}. The results of such processing can have implications on society and democracy that are far greater than matters of privacy, and therefore may require political and economic interventions beyond privacy engineering and associated technical solutions~\cite{KhanP19}.

%

\subsection{Steps forward}
It is probably not surprising that the gap between theory and practice raises many problems that seem intractable. The problems we encounter, however, also provide many possibilities both for research and change, some of which we discuss next.

\para{Structured agile}
There are already many corrections of agile practices that may provide opportunities for rethinking its activities with an eye on privacy engineering. The introduction of agile methods in organizations have not been without its problems. These include difficulties in planning and complexity as well as the cost of agile teams and use of third party services. The responses to these problems can be thought in terms of privacy design and engineering. For example:

\begin{itemize}[nosep]
\item Proposals for \emph{structured agile}: some actors have come to make proposals for more structured agile. Efforts like this may benefit from research into, for example, how much planning is needed to enable privacy engineering? 

\item Proposals for \emph{scaled agile}: Scaled agile are proposals for large companies that have existing structures that are siloed, exploring ways to introduce more coherence among functions and departments. With scaled agile, an organization coordinates between agile teams. Such coordination may help in capturing information flows, composition of privacy designs, enabling concerted privacy engineering practices, and also addressing questions around trust. These activities may also help implement an organization-wide privacy strategy.
\end{itemize}

%

\para{Standardization:} Standards, if done right, hold the potential to widely disseminate privacy engineering practices in the industry. 
The closing of the gaps is likely possible through standards that aspire to build a common vocabulary and to find ways to integrate knowledge from researchers, practitioners and legal experts~\cite{NISTIR8062}. 
A possible limitation is that standards need to remain abstract enough to offer flexibility of implementation across industries and practices. Every organization that obtains a standards certification tailors their practices to comply with the standards requirements. Thus, no two ways of complying with a standard are the same and standards have to allow specificities in the implementation by remaining at a higher abstraction level and introducing mainly principles to be followed.



\para{The limits and potentials of purpose limitation} One way to mitigate the many challenges that agile practices in service architectures pose to privacy design and methodologies is through regulation. Very much like proposals coming from privacy research, data protection laws promote minimizing collection of data and restricting its use to a specific purpose, although not necessarily technically. 

Limiting data collection to a purpose only contractually is not sufficient to curb the very broad data collection and dissemination practices that services architectures and agile development practices have normalized. 
In current software development practices, once data are collected for a given purpose, they can be fed into service architectures and used in agile processes as long as it is done transparently. To guarantee transparency, regulation promotes contractual obligations encoded in terms of service or privacy policies. In the end of the day, regulation does not encourage purpose limitation at the technical implementation and thus software developers rarely design software with this principle.
It is promising that regulators have recently started looking at privacy intrusive infrastructural designs from a purpose limitation perspective. This is the case in a recent inquiry into Google and IAB's Real-Time Bidding standard~\cite{Ryan18}, as well as the recent complaint that Microsoft ProPlus telemetry data violates data protection and in particular purpose limitation~\cite{privacyCompany}. 
The notion of purpose limitation has the potential to appeal to agile teams that produce services because the existing metaphors in the field are closely related.
Limiting the scope of what is done and how, through concepts such as bounded context and doing one thing well, is at the heart of software engineering practices. 
Introducing purpose limitation could potentially ripple throughout the end-to-end designs and tie the legal nature of purpose limitation to the technical implementation. 
How data protection laws and software engineering methods combined can better ensure data minimization is enforced in service ecosystems is an interesting point of future inquiry.

\para{Circling back to computer science and engineering}

The challenges above may determine the future of privacy designs and privacy engineering. In the meantime some low-hanging fruits can be addressed from a technical perspective:

\begin{enumerate}[nosep]
\item Empirical and Measurement Research: Capture in depth the challenges modularity, evolution, distributed trust and novel development methods raise for privacy designs and methodologies. Conduct empirical studies of current software engineering practices with an eye on privacy engineering, experiment with ways to introduce privacy designs and engineering methods. Expand work on web measurement from the current focus on advertisement to study software infrastructures.
\item Design for agile and services: Explore PETs that can be updated, that enable new encoding, or are composable. Study different models of PET composition using and going beyond service orchestration and choreography.
\item Support Developers: Consider ways in which current software engineering practices can be done in a privacy preserving manner: e.g., privacy preserving monitoring of service environment. Introduce testing suits and evaluation methods for (composition in) PETs. Study and mitigate privacy implications of developer activities like dark launches, parallel development, continuous integration. Engage in "brownfield" privacy engineering, i.e., retrofitting privacy into legacy systems in agile service environments.
\end{enumerate}

\Bt{
    New points to bring into the discussion:
    \begin{itemize}
        \item all these aspects that we evaluate the methodologies on (criteria from Table 2), are there methodologies that deal with them without the privacy aspect, and if yes, how? Is it possible to capture all these aspects within a methodology or is it a single-case treatment that we can hope for at best?
        \item 50\% of the PETs articles are on analysis and attacks -- could be of great value to package these as tests for new systems, or evaluation suites (are we vulnerable to known problems? extract the problems from the context, organize them in taxonomies of attacks and analysis, create standardized tests for new privacy systems)
        \item service composition layer as the almighty oracle --> even though everything is partitioned, in an enterprise context, there is usually/almost always a point where everything comes together (business transactions have to be verified, we can't expect that the payment of a bill will be actually received). However, data flows are of actual problem, as generating more data flows isn't a problem in services. Business rule engines, integration test suites, transactions, we always have a way to test the end result of an operation carried out in a system. Yet, distributed systems are notorious for their consistency issues (what do we do with the CAP theorem?) 
        \item microservices and choreography are snake oil, but the service-orientation isn't. Everyone might be talking how they do microservices but at the end of the day, it's only a monolithic microservices architecture. We can't let go of all control and the illusion of flexibility might come at a greater cost that the PR it generates (some trends are practiced so that companies can attract top talent or to satisfy market dynamics, not because of a real need) As long as we go beyond the hype and acknowledge that there is still control, we might be able to (1) study where who has control over what, (2) design systems that make use of the control mechanisms
    \end{itemize}
}

\bibliographystyle{plain}
\bibliography{thebibliography}{}








\end{document}